# IP BASED TRAFFIC RECOVERY: AN OPTIMAL APPROACH USING SDN APPLICATION FOR DATA CENTER NETWORK

## Ahmed Mateen[*]

## Qaiser Abbas[*]

**Abstract**

With the passage of time and indulgence in Information Technology, network management has proved its significance and has become one of the most important and challenging task in today's era of information flow. Communication networks implement a high level of sophistication in managing and flowing the data through secure channels, to make it almost impossible for data loss. That is why there are many proposed solution that are currently implemented in wide range of network-based applications like social networks and finance applications. The objective of this research paper is to propose a very reliable method of data flow: Choose best path for traffic using SDN application. This is an IP based method in which our SDN application implements provision of best possible path by filtering the requests on base of their IP origin. We'll distinguish among source and will provide the data flow with lowest traffic path, thus resulting in providing us minimum chances of data loss. A request to access our test application will be generated from host and request from each host will be distinguished by our SDN application that will get number of active users for all available servers and will redirect the request to server with minimum traffic load. It will also destroy sessions of inactive users, resulting in maintaining a best responsive channel for data flow.

**Keywords:Application Delivery Controllers (ADCs);Network Operating System (NOS);Equal-cost multi-path routing (ECMP);Application Interface (API)**

[*]**Computer Science Department, Agriculture University Faisalabad, Pakistan**

## 1. Introduction

Programming Defined Networks (SDNs) have been perceived as the cutting edge organizing worldview with the guarantee to drastically enhance arrange asset usage, improve organize administration; diminish working expense, and advance development and development. One of the primary thoughts to SDN isolate the information plan to control arrange by (i) removing the address to the equipment eg., Routers or Switches. Load adjusting innovation to premise on which today's Applications Deliveries Controllers work. In any case, the inescapability of load adjusting innovation does not mean it is universally comprehended, nor is it regularly considered from something besides an essential, organize driven perspective. To amplify its benefits, associations ought to comprehend both the nuts and bolts and subtleties of load adjusting. Stack adjusting got its begin as system based load adjusting equipment. It is the basic establishment on which Application constructed stack adjusting (taking after application based restrictive frameworks) appeared as system based machines. To provide the genuine establishing father of today's ADCs. Since those gadgets were application-nonpartisan & dwelled external of the application servers themselves, they could stack adjust utilizing direct system strategies. Fundamentally, these gadgets would show a "effective server" deliver the external planet, and clients endeavored to interface, they would further the association with the mainly suitable genuine server doing bi-directional system deal with interpretation (NAT). The most conveyed SDN convention, OpenFlow [5][6], empowers the correspondence among the information level surface and a distant controller, where together control flat and applications live. Each OpenFlow-consistent switch has a flow table contain an arrangement of tenets (i.e., flow-table passages) appropriate for flows taking care of, where a flow is planned as a grouping of bundles allocation some title fields esteem. Every bundle got by the switch is thought about next to the flow-table. In the event that a coordinating section is create, the linked activity (e.g., onward out a specified port or drop) is perform and a few counter (e.g., bundles/bytes Tx/Rx) are overhauled. Something else, the parcel is send to the organizer, which sets up how to switch it and upgrades the tenets in the switches flow-table. Versatility is an all around satisfied necessity in IP/MPLS systems [7], yet it is an open issue for SDNs. In fact, the programmed reconfigurable of traffic ways in the event of connections or hubs disappointments is not any more accessible, because switches flow-table passages are currently to define and overhauled just by the organizer. As a matter of course, upon disappointment events the controller does not play out any activity. Our paper goes for conquering this issue by presenting another SDN control application, created on top of POX [9]. In addition, to consider the distinctive necessities of traffic flows as far as recuperation time, our controller underpins Restoration and also Protection.

Security includes the calculation and setup of the OpenFlow policy for mutually WP and RP in the meantime. In our engineering two diverse defense systems are accessible: "devoted assurance" and "on-the-fly security". In the first case equally WP and RP are at the same time actuated and client traffic is duplicated on them (1+1protection), with no administration blackout if there should be an occurrence of disappointments. Rather, in the second box the RP is initiated by the organizer through the evacuation of the sending policy for the WP simply past a disappointment is distinguished. Programming Defined Networking (SDN) has the possible to empower advancement in the improvement of PC frameworks. It depends on the advantage of detaching the control and data planes. In out of date frameworks, switches, switches, firewalls, action distributer are particular dealer exact gear gadgets having unequivocally connected data and control plane. The Functionality of these devices cannot be changed logically. In SDN, establishment layers are humble thing silicon devices. data plane. On top of control plane, framework applications, for instance, trading, coordinating, firewall and development distributer can be executed by using mix of stream. Controller self-administering sets up every stream in per stream based Routing.

1. Support software: Hardware maker offer a few utilities with the assistance of compiler and an assortment of bolster programming from the working framework are the fundamental dependence of programming characterizing system for advancement of use. Issues that can be created by the movement of information will be huge than the system limit.

2. Software Defining Network: With the correlation of present web activity and earlier day movement stacking structure to characterize the systems of old web movement dispersion and new patterns of activity load to appropriate and conquer the issue of load adjusting.

3. Network operating System: In this figure the system working framework is play out the fundamental action of this figure arranged working framework to play out the heap adjusting action to rise to appropriate the activity.

In this paper, to propose an answer for handle clog of information activity utilizing instrument in view of our already way IP based application to tally the guests and to choose the best for lose free and fast reaction [8]. This joined arrangement ought to have the capacity to ensure the system execution as far as misfortune free casing conveyance and high throughput. This paper is sorted out as takes after: Section II briefly audits some related work. Segment III depicts the real systems of IP based identification. Segment IV exhibits and examines the execution of this application. A brief conclusion is attracted Section V took after by some discourse about the proceeding with work.

## 2. Load balancing basics

With this basic vocabulary built up, how about we look at the essential load adjusting exchange. As delineated, the heap balancer will regularly sit in-line between the customer in load adjusting, this is not an administer, but rather to a greater extent a best practice in a run of the mill organization. How about the likewise accept that the heap balancer is as of now designed with a external server that focuses to a bunch comprising of two administration focuses. In this organization situation, it is basic for the hosts to have an arrival course that focuses back to the heap balancer with the goal that arrival movement will be prepared during it on its way reverse to the customer.

This extremely straightforward case is generally direct, however there are two or three key components to observe. To start with, to the extent the customer knows, it sends parcels to the virtual server and the virtual server reacts—basic. Second, the NAT happens. This is the place the heap balancer replaces the goal IP sent by the customer (of the virtual server) with the goal IP of the host to which it has stacked adjust the demand. footstep three is the second 50% of this procedure (the division that make the NAT "bi-directional").Rather, the heap balancer, recalling the association, modifies the bundle so that the basis IP is that of the virtual server, therefore taking care of this issue.

## 3. Related work

SDN (software-defined networking) is a technology in the field of computer networking which is in the blink of an eye creating critical intrigue. It started from a venture that started at UC

Berkeley and Stanford University around 2008 [4]. SDN is at present observed as one of the developing ways to deal with PC organizing that permits arrange scientists to oversee organize benefits through reflection of lower level usefulness [5, 6]. This is accomplished by decoupling the system control that settles on choices about where activity is sent (the control plane) from the sending frameworks that forward movement to the chose goal (the information plane). The system turns out to be straightforwardly programmable and permits the foundation to be preoccupied for applications and system administrations. The specialists and merchants of these frameworks assert this improves organizing [7]. At its center, SDN offers higher adaptability and quick steering of activity streams. Inside the structure of this division, engineers can use the control plane to change the conduct of the system without physical adjustment of the current system foundation execution. This permits engineers to direct investigations adaptable and effectively and empowers the quick sending of new system designs. Inside the SDN engineering, the application layer gives clients an extensive variety of inventive administrations and applications, while the control layer is accomplished by SDN programming on the server. For usability, the SDN programming incorporates a uniform application program interface (API) [8]. The information layer is included bland system gadgets which can give equipment or exchanging operations which are programming characterized in the control layer and imparted through the OpenFlow standard protocol [9].

The OpenFlow protocol is a fundamental element for building SDN solutions. It is the first standard communication protocol defined between the control layer and the infrastructure layer in SDN architecture [10, 11]. OpenFlow utilizes the idea of streams to recognize arrange activity in light of coordinating guidelines that can be statically or powerfully customized by the SDN control programming. Switches are in charge of applying the best possible activities on parcels and overhauling records in the stream table section. The switches essentially forward bundles as indicated by the pertinent section in the stream tables without being worried with how to build or change the stream table. The controller makes and introduces a govern in the stream table for the comparing parcel if important, and the controller may whenever deal with all switches by the stream table. OpenFlow-based SDN designs give to a great degree granular control, empowering the system to respond to continuous changes from the application or the administration client [12]. OpenFlow-based SDN advancements increment the transmission capacity ability, dynamic nature of uses and fundamentally diminish operation and administration intricacy [13].

At present, the existing traffic scheduling algorithms mainly include Round-Robin scheduling algorithm and Greedy scheduling algorithm. These scheduling algorithms have some downsides, for example, high cost, low unwavering quality, and low adaptability. The capacity of information calculations to manage mass-activity turns out to be more essential with the expansion in versatile client. To tackle the intricate choice issue that system confronts, likelihood determination calculation can be viewed as a sort of good strategy. For likelihood determination calculation, the worries are not a matter of a flag decision but rather the creating pattern of server movement and the heap of servers. Quickly, in the part of the arrangement space, to get the presence of the ideal arrangement under complex environment. In every emphasis, the spare an arrangement of competitor arrangements and pick better attainable arrangement by utilizing likelihood choice calculation in light of the mapping of server load and afterward deliver another era of applicant arrangements. The procedure is rehashed until F-test esteem focalizes to the limit.

### 3.1. Preliminary of segment routing

For every couple of host interchanges inside the similar SDN/SR space, a system use an inside door convention (IGP, for example, the release most brief way first (OSPF) convention, to empower steering to the goal as a matter of course. SR [6] utilizes a hub fragment to rep-detest the activity that a bundle ought to take after to take the briefest course. At the end of the day, each hub in a similar SR area keeps a hub section data connection to each of alternate hubs in its sending table, as the defaulting policy and the hub portion rep-despises worldwide mindfulness. What's more, SR receives a nearness segment to control movement, which speaks to the activity that the parcel ought to exchange to a particular departure information interface with a contiguous hub, and this contiguousness fragment speaks to nearby mindfulness. In any case, not at all like in the hub portion, every hub just needs to introduce its local nearness section leads in its sending table. The blend of hub fragments and nearness portions shapes an arrangement rundown of marks that are connected to the parcel title by the SDN organizer, and they are reflect in a split second as the coveted activity way. Along these lines, SR brings a few requests of scaling increases, as it doesn't hold any state for the streams in the middle gadgets. Be that as it may, SR adds to another recognizable issue, since it utilizes a MPLS mark field to put the section names. Accordingly, SR may require a bigger bundle header, which decreases the accessible band-width. The following paragraphs summarize a portion of the past reviews identified with the SDN innovation. To start with, the creators in [14] believe the issue of deciding the ideal parameter for SR in disconnected and online bugs. The creators suggest a movement lattice neglectful calculation for the disconnected case, and a new calculation for the online issue. In the online issue, the system has a focalized organizer that can utilize an online way to deal with tackle the SR issue, as is done in our SDN-based situation. These creators give recipes and straight projects for characterizing the ideal parameters of SR. Their paper concentrates on deciding the ideal parameters as movement rip qualities. These qualities are connected to reduce the most pessimistic scenario interface usage by considering rise to cost multi-way directing (ECMP) in disconnected cases. The activity rip values additionally serve to minimize dismissals of solicitations in online issue. Be that as it may, our examination is devoted to outlining a productive defeating calculation for better system execution, (for example, enhanced system results and dismissal rate) and to concentrate on discovery the single most limited way without ECMP. Think the connection criticality and the connection lingering transmission capacity in assessing the heaviness of a connection. Additionally, our calculation restrains the maximal way length of a course to lessen the system's utilization of data transfer capacity.

The authors in [15] expressed the cases of vitality utilization while conveying substantial scale appropriated foundations. They propose the utilization of SR-based vitality proficient movement building to diminish the vitality utilization of spine systems. Through the SDN approach, the system can specifically switch of subset of connections. The creators actualize this strategy in the OMNET ++ test system that progressively chooses the quantity of force on connections, and accordingly spares vitality. The objective of these creators is to decide the status of system gadgets, paying little mind to whether they are required to exchange information. Our paper, nonetheless, means to construct a transmission capacity fulfilling way that builds the aggregate system throughput and lessens the demand dismissal rate, as opposed to diminishing the vitality utilization.

The authors in [16] present a design that coordinates the SDN worldview with SR-based movement building. Their paper concentrates on the issue of mapping registered ways onto SR ways. They propose a SR way task calculation that means to locate the briefest rundown of portions relating to the fancied way. Our work, not withstanding, concentrates on the issue of a steering calculation as opposed to a SR mark task.

The authors in [17] explained a subdivision list-encoding algorithm to convey a given way, which minimize the fragment list profundity in SR-based systems. Also, these creators give a technique to ECMP-mindful most limited way calculation that is liable to a measured arrangement of numerous imperatives. Our job likewise takes the mark heap profundity into thought, and proposed a directing calculation to diminish the additional cost of the bundle header, which is brought about by name heap profundity. Notwithstanding the mark stack profundity issue, consider the adjust of activity load in our directing calculation, with a plan to enhance execution as far as the system through-put and the dismissal rate. Subsequently, our work concentrates on the issue of enhancing the steering calculation as opposed to taking care of the portion list calculation issue.

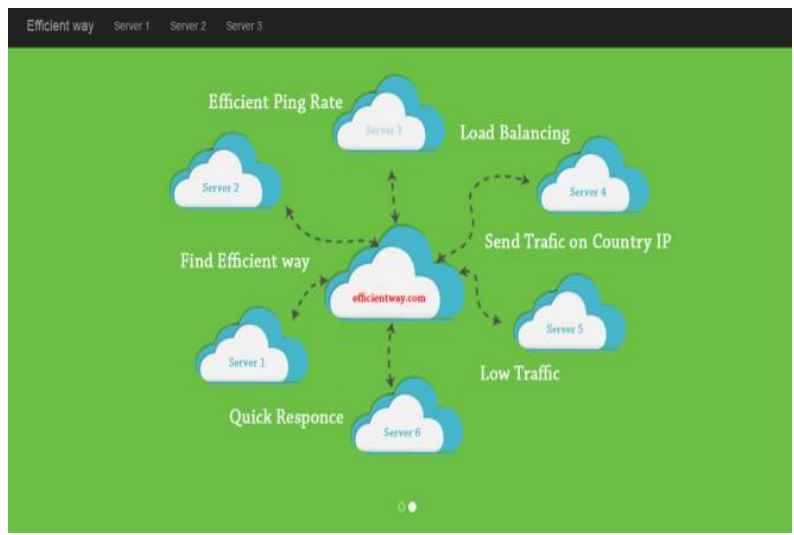

Figure 1.*Design of Efficient.com*

3.2. Traffic engineering

Many existing works regard activity designing as an in unnecessary device for redesigning system execution, which workings by expressly coordinating the movement throughout a limited, focused system asset. Movement designing arranges the directing plan to control how activity is steered over the system, to streamline organize execution and utilize organize assets effectively. It is important that movement designing techniques use the deliberate activity lattice for the determination and administration of system clog. The movement grid speaks to the volume of activity between sets of source-and-goal matches over a given time interim, and this component is a key calculate arrange arranging. The estimation of the movement lattice includes all the while gathering activity stream estimations and data on steering. SDN permits a more dynamic system estimation, which can all the more effectively decide the precise and auspicious movement grid because of OpenFlow and the unified controller. Subsequently, when th think about movement building, we can procure expectations of future activity inclines through the evaluated activity grid, and it will b discover great directing setups.

## 4. Design and Implementation of Our Scheme

Load balancing provides a straightforward approach to build the transfer speed of servers and other system gadgets and upgrade information parcel preparing limit and system throughput to at last enhance the ease of use and adaptability of a system [14, 15]. Stack adjusting means to improve asset utilize, amplify throughput, minimize reaction time, and abstain from over-burdening any single asset. The significance of server load adjusting is perceived with the end goal that techniques to enhance stack adjusting are effectively and consistently explored. In contrast with the fast improvement of system innovation, the development rate of server processor speed and memory get to is nearly moderate. At present, the handling overhead of servers is a noteworthy bottleneck of the system advancement. Incomprehensibly, with the improvement of rapid systems and expanding requests for administrations, numerous venture server farms and entrance servers are getting to be overpowered by the dangerous development in information movement. Stack adjusting is the key innovation used to convey information requests over a group of server frameworks.

In this scheme of server load balancing based of a forward exchanging strategy, a novel technique is proposed in this paper by using the Network Address Translation (NAT) in the SDN design to develop a half breed stack adjusting model. NAT alludes to utilizing a virtual deliver to speak to the genuine server address and reworking the goal address of the demand bundle. Eventually, information retransmission is performed [16, 17]. The present load adjusting strategies are portrayed by high speculation, high utilization, low dexterity, and low dependability. A large number of these issues can be understood by programming characterized organizing. This paper presents another likelihood strategy for load adjusting in view of the difference examination in SDN systems.

## 4.1. Server load balancing

This application working as a load balancing, at the point when somebody send a ping request to my server it check stack shape all other server and send it to the minimum occupied server, All server are working parallel, behind the scene I am checking the no of guest on every server and send the demand to the slightest occupied server, they check it for all the time then it distinguishes the server my application to choose the misfortune free ideal way and send the activity.

## 4.2. Node, Host, Member, and cluster Server

Most stack balancers have the idea of a hub, host, part, or server; several have each of the four, however they denote diverse stuff. There are two essential ideas that they all attempt to state. One idea as a rule call a hub or server is the possibility of the corporal server itself that will get movement from the heap balancer. This is identical with the IP address of the physical server and, without a heap balancer, For whatever is left of this paper, it will suggest this thought as the host.

The second idea is a part (some of the time, lamentably, likewise called a hub by a couple of makers). A section is for the most part to some degree more portrayed than a server/center in that it fuses the TCP port of the honest to goodness application that will get action.

## 4.3. Virtual Server

Although not always the issue, today there is little dispute about the term virtual server, or virtual. Note that like the meaning of administrations, virtual server for the most part incorporates the application port was well as the IP address. The expression "virtual administration" would be more with regards to the IP: Port tradition; but since most merchants utilize virtual server, this paper will keep utilizing virtual server also.

In this figure to describe the efficient.com application t o distributes the traffic on server 1 is lightweight traffic at this time and accept the traffic load.

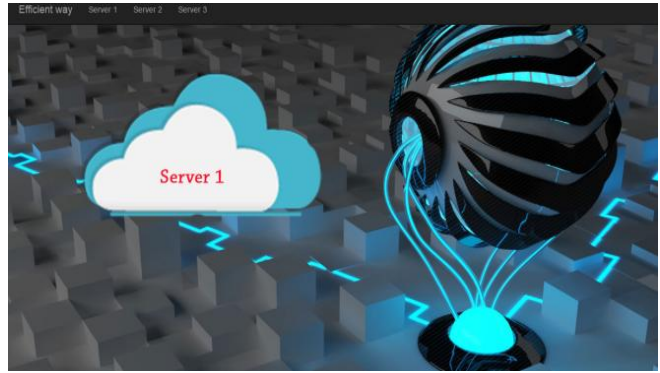

Figure 2.*Server 1 Distribute traffic is figure*

## 5. Variance Analysis

### 5.1. F -Test

Analysis of variance (ANOVA) is a set of statistical models which are utilized to analyze down the difference between group implies and their related methodology, create. In the ANOVA setting, the observed difference of a specific variable is apportioned into components attributable from various sources of variety. This paper utilizes a variance analysis method to determine whether the averages of several sets of data are equal by analyzing data statistics.

For analyzing the statistical characteristics of port flux, this paper adopts the t-test technique to identify whether there are noteworthy contrasts among ports keeping in mind the end goal to figure out whether the operation is legitimate for the present state. Also, in light of the fact that information stream in the system is arbitrarily chossen, the activity from every port can be seen as free with a typical circulation. The general contrasts are isolated into two essential classes of inside gathering variety and between-gathering variety. [18] Contrasts in the between-gathering class are ascertained to assess a significant scattering between the normal estimations of intra group movement and the populace mean. Contrasts in the inside gathering class are ascertained to assess the scattering between an impartial specimen in a similar gathering and the populace mean. F- test examination is a factual procedure that is utilized to recognize an arrangement of gatherings in light of contrasts. The mean square is acquired through the estimation of contrasts between the two sections separated by their degrees of flexibility. The F- investigation esteem is characterized as the proportion of the "intra-" and "between " contrasts, as indicated by the near examination of the - assessment and centrality level limit This will figure out if there is a huge contrast between ports.

$$F = \frac{MS_b}{MS_w} = F(df_b, df_w).$$ (1)

Equation No:1 [18] F means the "*F Statistic*."

Based on the above conclusions, the F -test formula.

In (1), MSb is the between-group difference, MSw is the within-group difference. To clarify by example, let A1, A2,……, AK be a factor set having K different parts, let $n_i$ be the number of monitoring times at level Ai , let  be related to the traffic, and let Xi1 , Xi2,.....,Xin be the set of X samples at level Ai. Consider

$$\overline{X} = \frac{1}{N} \sum_{i=1}^{k} \sum_{j=1}^{n_i} X_{ij}, \quad i = 1, 2, 3, \ldots, k,$$

Equation No: 2 [19] X- bar or "Standard *Deviation*."

In (2), X –bar equals the average of all of the traffic values. K refers to the number of groups.  N refers to the number of the total monitoring times. Consider

$$SS_t = \sum_{i=1}^{n} \sum_{j=1}^{K} \left( X_{ij} - \overline{X} \right)^2.$$ (3)

Equation No: 3 [20] "*Total Sum of Squares.*" between the n data points and grand mean. Then Calculate

SSt= SSb + SSw

In (3), SSt is the total sum of squares, which equals a square sum of deviations between every sub sample in population and population mean. Consider

$$SS_b = \sum_{j=1}^{k} n_j \left( \overline{X_j} - \overline{X} \right)^2.$$ (4)

Equation No: 4 [20] "*Sum of Squares between-Sample*"

In ([4](4)), SSb is between-group sum of squares, which refers to a sum of squares of the deviations about the value between each group mean and population mean. Consider

$$SS_w = \sum_{i=1}^{n} \sum_{j=1}^{k} \left( X_{ij} - \overline{X} \right)^2. \tag{5}$$

Equation No: 5 [20] "*Sum of Squares for Within Sample.*"

In ([5](5)), SSw is within-group sum of squares, being equal to a sum of squares of the deviations about the value between every subsample value in group and each group mean. Consider

$$MS_b = \frac{SS_b}{df_b} = \frac{n_1 \left( \overline{X_1} - \overline{X} \right)^2 + n_2 \left( \overline{X_2} - \overline{X} \right)^2 + \cdots + n_k \left( \overline{X_k} - \overline{X} \right)^2}{N-1} = \frac{\sum_{j=1}^{k} n_j \left( \overline{X_j} - \overline{X} \right)^2}{N-1}, \tag{6}$$

Equation No: 6 [18] "*Mean Squares between.*" divided by the between group degree of freedom.

$$MS_w = \frac{SS_w}{df_w} = (n_1 - 1)S_1^2 + (n_2 - 1)S_2^2 + \cdots + (n_k - 1)S_k^2 = \frac{\sum_{i=1}^{n} \sum_{j=1}^{k} \left( X_{ij} - \overline{X} \right)^2}{N-K}. \tag{7}$$

Equation No: 7 [18] *"Mean Squares within or Error."* divided by error degree of freedom.

In ([6](6)), the division of by the degree of freedom dfb returns a numeric result and assigns the result to MSb which indicates within-group variance. Similarly, MSw which refers to between-group variance can be obtained according to the result of ([7](7)). [18] The F -test is employed to compare the factor of the total deviations. The F -inspection value is defined as the ratio of the "intra-" and "inter-" differences. An observed value F of which is greater than the critical value of F determined from tables indicates that there are significant differences among groups. Conversely, a small F-test value which does not exceed the critical value of F determined from tables indicates that there is no fundamental distinction among groups.

## 5.2. T-Test and Multiple Comparisons

Based on the results of the above calculations, we acquire the F - review esteem which must be utilized to demonstrate whether there are noteworthy contrasts among gatherings. The F - investigation esteem does not make it clear which of these gatherings, which ought to be very few, contain essential contrasts. [18] There is a need to analyze the ascertained midpoints encourage by receiving the numerous T-tests. Before talking about the different T-tests, we first concentrate on two free T-tests and accept that. The T-test method expression is shown as follows:

$$t^2 = \frac{\left( \overline{X_0} - \overline{X_1} \right)^2}{SS_w^2 \left( 1/n_0 + 1/n_1 \right)} = \frac{n_0 + n_1}{n_0 n_1} \frac{\left( \overline{X_0} - \overline{X_1} \right)^2}{SS_w^2}. \tag{8}$$

Equation No: 8 [22] "*T squares of Test*." To used for multiple domain.

Then

$$t = \sqrt{\frac{n_0 + n_1}{n_0 n_1}} \frac{|\overline{X_0} - \overline{X_1}|}{\mathrm{SS}_w}. \qquad (9)$$

Equation No: 9 [21] *"Sum of total No"*

Let, n =n0+n1 and n denotes the sum of monitoring numbers According to the above principle, there is a formula of multiple t-test about k (k>2) ports. As H0 is true, the hypothesis is as follows:

$$H_0: \mu_1 = \mu_2 = \cdots = \mu_k, \, H_1: \mu_i \, (i = 1, \ldots, k) \text{ are not all equal}$$

Then the multiple T-test method is with the formula as follows:

$$t_\alpha (\mathrm{df}_w) = \frac{|\overline{X_i} - \overline{X_j}|}{\sqrt{2\mathrm{MS}_w/n}}. \qquad (10)$$

*Equation No:10 [21] "Least Significant Difference (LSD)."*

where, xi . xj are the point of any two of these averages; MSw is the mean square and dfw is the degree of freedom. Consider

$$D_\alpha (\mathrm{df}) = t_\alpha (\mathrm{df}_w) \sqrt{\frac{2\mathrm{MS}_w}{n}}. \qquad (11)$$

Equation No: 8 [21] *"Least Significant Difference (LSD)."*

That is, if the difference between any two averages reaches or exceeds the significance level, then the null hypothesis is rejected. It is then necessary to proceed effectively with dynamic load balancing to avoid contention.

## 6. Experiment Result and Performance Analysis

In the examination, our working framework was Ubuntu 14.04.3-desktop-amd64, controller was Floodlight adaptation 1.0, Mininet 2.1.0 which is a system emulator for the making of virtual system utilizing the Ubuntu piece was utilized to characterize the topology of the entire system, and Open vSwitch 2.3.2 was utilized to mimic the required OpenFlow switch. With a specific end goal to quantify the trial, an OpenFlow test stage was worked as portrayed in Figure 3. The SDN organize demonstrate contains four free server hubs, three OpenFlow switches, and a Floodlight controller. Our examination exhibited some server code written in Python, however practically a similar outline would apply for almost any dialect. Python accompanied a basic stage worked in HTTP server.

In the test, Mininet was to first form engineering with various ways. Every server spoke to a physical machine in Mininet and had its own real IP address. We made a virtual IP address which is publicized from the NAT, and approaching movement bound to this virtual IP address was steered by Floodlight controller to various genuine IP addresses. Made robotized scripts of get to demands on the customers. Next, To provide constant movement flux factual examination through the investigation of difference and activity booking calculation modules which were incorporated into the Floodlight controller and picked suitable way in light of our investigation result.

These switches don't confine the transmission speed and work with greatest connection rate. Java code was utilized to actualize the investigation of change and movement booking calculation modules which are incorporated into the Floodlight controller.

By examining the aftereffects of the analysis and contrasting and different calculations, the change in the execution of the proposed calculation was checked. The investigation comprises of instatement of information stream, movement examination of change, and calling of the heap adjusting calculation.

Figure 6Shows bends for all planning calculations in one figure and their association with the F-test esteem. The littler the estimations of the F - test in a progression of analyses are, the less the distinctions among the observing information activity of every port are. As appeared here, the proposed calculation coordinates well with the Greedy calculation. With the expansion of the seasons of collaboration transmission, the F - test qualities are apparently decreased, and these strategies can change all servers stack adjusting precisely continuously and divert movement all the more effectively. It is likewise clear from the assume that the Round-Robin calculation is not the best one which implies there are extraordinary contrasts among servers' heap, and this technique is irregular.

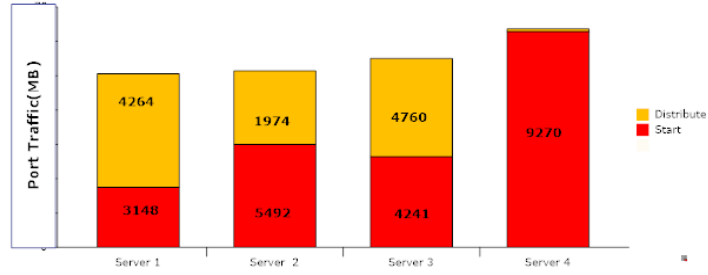

Figure 3. *IP based Application Working*

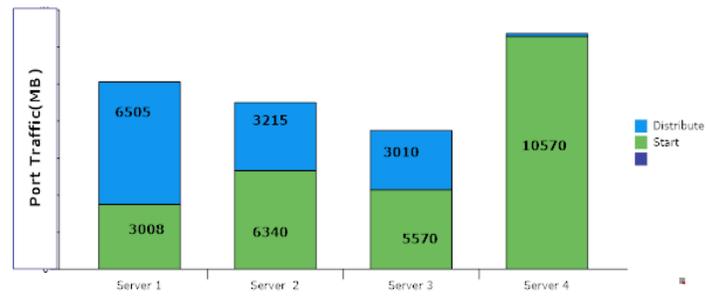

Figure 4. *Greedy Algorithm*

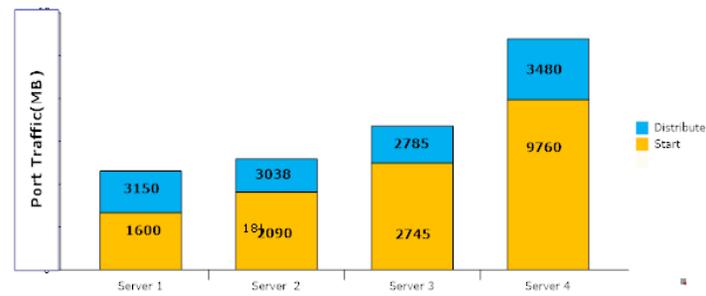

Figure 5. *R. Robim Algorithm*

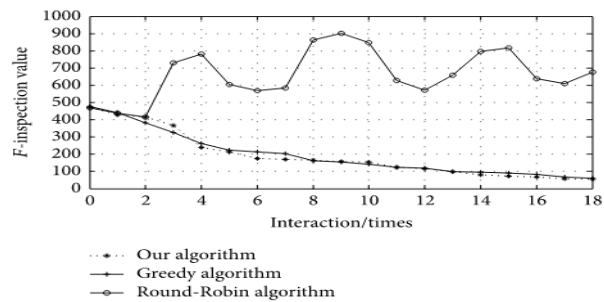

Figure 6.*Final Results*

The examination requires an estimation of quantitative investigation in view of an estimation server in the SDN. The controller can spare information plane data and interchange relationship amongst controller and OpenFlow change to a neighborhood log record. The estimation server (testing server) executes synchronization operation on the Floodlight controller and servers and gets the movement information. The general plan of the testing stage is appeared.

Through a simulation test, the execution of the proposed calculation is checked. Accept that a client demands access to a virtual address. The correspondence time of every server including a web benefit demand is 5 seconds.

At the point when unnecessary server stack happens on one server, the accompanying three load adjusting calculations are executed separately: Round-Robin planning, Greedy booking, and likelihood planning. The activity at every server is caught by the change investigation module in the Floodlight controller and this bit of data is spared to log records. Figures 3, 4, and 5 represent the beginning and completion places of port activity. Figures 3 and 4 demonstrate the reaction of the proposed calculation and the Greedy calculation, individually. It can be seen from these assumes that the pinnacle estimations of the considerable number of segments are basically level, implying that every one of the four servers can stack adjust utilizing by the two strategies. In a genuine domain, it is not important to keep strict harmony whenever. Figure 5 demonstrates the aftereffect of the Round-Robin booking calculation. There is impressive contrast in the pinnacle estimations of every section, which shows that this strategy is incapable.

## 7. Conclusion

It is central to appreciate that basic load balancing technology, while still being used, is presently just measured an element of Application Delivery Controllers. ADCs advanced from the principal stack balancers and finished the administration logically handle; they can enhance accessibility, as well as influence the defense and execution of the application administrations being asked. Today, most associations understand that basically having the capacity to achieve an application doesn't make it utilizable; and unusable applications mean squandered time and cash for the undertaking sending them. ADCs empower associations to solidify network based administrations like SSL/TLS pass on, storing, pressure, rate-molding, interruption recognition, application firewalls, and even out-of-the-way access into a solitary tip that can be public and reused over all application administrations and all hosts to make a virtualized Application Delivery Network. Essential load adjusting is the establishment without which none of the improved usefulness of ADCs would be possible.